\documentclass{article}[12pt]
\usepackage{cleveref}
\usepackage{color}
\usepackage{graphicx,amssymb,amsfonts,amsmath,amssymb,amscd,amstext,mathrsfs}

\pdfoutput=1
\def\be{\begin{eqnarray}}
\def\ee{\end{eqnarray}}

\def\Vol{\operatorname{Vol}}

\begin{document}
\thispagestyle{empty}

\begin{flushright}
YITP-SB-14-48
\end{flushright}

\begin{center}
\vspace{1cm} { \LARGE {\bf 
Thermal Corrections to R\'enyi Entropies for Conformal Field Theories
}}
\vspace{1.1cm}

Christopher P. Herzog and Jun Nian \\
\vspace{0.8cm}
{ \it C.~N.~Yang Institute for Theoretical Physics, 
Department of Physics and Astronomy \\
Stony Brook University, Stony Brook, NY  11794}

\vspace{0.8cm}

\end{center}

\begin{abstract}
\noindent
We compute thermal corrections to R\'enyi entropies of $d$ dimensional conformal field theories on spheres.  
Consider the $n$th R\'enyi entropy for a cap of opening angle $2 \theta$ on  $S^{d-1}$.  
From a Boltzmann sum decomposition and the operator-state correspondence, the leading correction is related to a certain two-point correlation function of the operator (not equal to the identity) with smallest scaling dimension.  
More specifically, 
via a conformal map, the correction can be expressed in terms of the two-point function 
on a certain conical space with opening angle $2\pi n$.  
In the case of free conformal field theories, this two-point function can be computed explicitly using the method of images.  
We perform the computation for the conformally coupled scalar.
From the $n \to 1$ limit of our results, we extract the leading thermal correction to the entanglement entropy, reproducing results of
arXiv:1407.1358. 

%
%We start with the two-point function on an $n$-fold covering of the sphere $S^d$ in $\mathbb{R} \times S^d$. 
%From it we calculate the thermal correction to the R\'enyi entropy for a cap-like region of opening angle $\theta$ 
%on $S^d$. In some appropriate limits, our results reproduce the analytical results in Refs.~\cite{CardyHerzog, Herzog}. 
%We also compare our results with the conformally coupled scalars on the sphere numerically, and find a consistent match.

\end{abstract}

\pagebreak
\setcounter{page}{1}

\section{Introduction}

Entanglement entropy plays an increasingly important role in different branches of physics.  Proposed as a useful measure of the quantum entanglement of a system with its environment, entanglement entropy now features in discussions of black hole physics
\cite{Bombelli:1986rw,Srednicki:1993im}, renormalization group flow \cite{Casini:2006es,Casini:2012ei}, and quantum phase transitions \cite{OsborneNielsen,Vidal:2002rm}.  A closely related set of quantities are the R\'enyi entropies.
In this paper, our modest goal is to obtain thermal corrections to R\'enyi entropies for conformal field theory (CFT).

We adopt the conventional definition of entanglement and R\'enyi entropy in this paper. Suppose the space on which the theory is defined can be divided into a piece $A$ and its complement $\bar{A}=B$, and correspondingly the Hilbert space factorizes into a tensor product. The density matrix over the whole Hilbert space is $\rho$; 
then the reduced density matrix is defined as
\be
  \rho_A \equiv \textrm{tr}_B \rho\, .
\ee
The entanglement entropy is the von Neumann entropy of $\rho_A$,
\be
  S_E \equiv - \textrm{tr} \rho_A \, \textrm{log} \, \rho_A\, ,
\ee
while the R\'enyi entropies are defined to be 
%To calculate the entanglement entropy, one usually first calculates a related quantity, 
%which is called the R\'enyi entropy. Its definition is
\be
  S_n \equiv \frac{1}{1-n} \textrm{log}\, \textrm{tr} (\rho_A)^n\, .
\ee
Assuming a satisfactory analytic continuation of $S_n$ can be obtained, the entanglement entropy can alternately be expressed 
as a limit of the R\'enyi entropies:
%The relation between the R\'enyi entropy and the entanglement entropy is
\be
  \lim_{n\to 1} S_n = S_E\, .
  \label{analyticcont}
\ee

To apply the results to a real system, it would be useful to know the thermal corrections to the entanglement entropy $S_E$ and the R\'enyi entropy $S_n$. Ref.~\cite{CardyHerzog} found universal thermal corrections to both $S_E$ and $S_n$ 
for a CFT on $S^1 \times S^1$.  
The CFT is assumed to be gapped by having placed it on a spatial circle of circumference $L$, 
while the circumference of the second circle is the inverse temperature $\beta$. The results are
\begin{align}
  \delta S_n & \equiv  S_n(T) - S_n(0) = \frac{g}{1-n} \left[\frac{1}{n^{2\Delta - 1}} \frac{\textrm{sin}^{2 \Delta} \left(\frac{\pi \ell}{L} \right)}{\textrm{sin}^{2 \Delta} \left(\frac{\pi \ell}{n L} \right)} - n\right] e^{-2 \pi \beta \Delta / L} + o \left(e^{-2 \pi \beta \Delta / L}\right)\, ,\\
  \delta S_E & \equiv S_E(T) - S_E(0) = 2 g \Delta \left[1 - \frac{\pi \ell}{L}\, \textrm{cot} \left(\frac{\pi \ell}{L} \right) \right] e^{-2 \pi \beta \Delta / L} + o \left(e^{-2 \pi \beta \Delta / L}\right)\, ,
\end{align}
where $\Delta$ is the smallest scaling dimension among the set of operators not equal to the identity and $g$ is their degeneracy.  
The quantity $\ell$ is the length of the interval $A$.\footnote{%
 The fact that $S_E(T) - S_E(0) \sim e^{-2 \pi \beta \Delta / L}$ is Boltzmann suppressed was conjectured more generally for gapped theories in Ref.\ \cite{Herzog:2012bw}.
 That $S_E(T) - S_E(0)$ might have a universal form for 1+1 dimensional CFTs was suggested by the specific examples worked out in Refs.\ \cite{Azeyanagi:2007bj,Herzog:2013py,Barrella:2013wja,Datta:2013hba}.
 See Ref.~\cite{Chen} for higher order temperature corrections when the first excited state is created by the stress tensor.
 %calculated the R\'enyi entropy of one single interval on a circle at finite temperature 
 %to a few higher orders both in conformal field theory and in a holographic way.

}

To generalize the results of Ref.~\cite{CardyHerzog} to higher dimensions, Ref.~\cite{Herzog} considered thermal corrections to the entanglement entropy $S_E$ on spheres. More precisely, a conformal field theory on $S^1\times S^{d-1}$ is considered in Ref.~\cite{Herzog}, where the radius of $S^1$ and the one of $S^{d-1}$ are $\beta/2 \pi$ and $R$ respectively. The region $A \subset S^{d-1}$ is chosen to be a cap with polar angle $\theta < \theta_0$.  Then the thermal correction to the entanglement entropy $S_E$ is
\be
  \delta S_E = g \Delta\, I_d(\theta_0)\, e^{-\beta \Delta / R} + o \left(e^{-\beta \Delta / R}\right)\, ,
  \label{deltaSE}
\ee
where
\be
\label{Idintegral}
  I_d (\theta_0) \equiv 2 \pi \frac{\textrm{Vol} (S^{d-2})}{\textrm{Vol} (S^{d-1})} \int_0^{\theta_0} d\theta\, \frac{\textrm{cos} \theta - \textrm{cos} \theta_0}{\textrm{sin} \theta_0}\, \textrm{sin}^{d-2} \theta\, .
\ee
Ref.\ \cite{Herzog} noticed that this result is sensitive to boundary terms in the action.  For a conformally coupled scalar, these boundary terms mean that the correction to entanglement entropy is given not by Eq.~\eqref{deltaSE} but by Eq.~\eqref{deltaSE} where $I_d(\theta_0)$ is replaced by $I_{d-2}(\theta_0)$.  
%For free fermions, however, there will be no such shift.

A natural question is how to calculate the thermal corrections to the R\'enyi entropy in higher dimensions. We would like to address this issue in the paper. Our main results are the following. The thermal correction to the R\'enyi entropy for a cap-like region with opening angle $2 \theta_0$ on the sphere $S^{d-1}$ in $\mathbb{R}\times S^{d-1}$ is given by
\be
\label{mainresult}
  \delta S_n = \frac{ n}{1-n}  \left( \frac{\langle \psi(z)\, \psi(z') \rangle_n}{\langle \psi(z)\, \psi(z') \rangle_1}- 1 \right) e^{-\beta E_\psi} + o \left(e^{- \beta E_\psi}\right)\, ,
\ee
%\be
%  \delta S_n = \frac{g\, n}{1-n}  \left[\left(\frac{\langle \phi(y)\, \phi(y') \rangle_n}{\langle \phi(y)\, \phi(y') \rangle_1}\right)^{\Delta / \Delta_0} - 1 \right] e^{-\beta E_\psi} + o \left(e^{-2 \beta E_\psi}\right)\, ,
%\ee
where $\psi(z)$ is the operator that creates the first excited state of the CFT and $E_\psi$ is its energy.\footnote{% 
 For simplicity, we have assumed that the first excited state is unique.  For a degenerate first excited state, see the next section.
}
If we assume that $\psi(z)$ has scaling dimension $\Delta$, then we know further that $E_\psi = \Delta / R$.  
The two point function $\langle \psi(z)\, \psi(z') \rangle_n$ is evaluated on an $n$-fold cover of ${\mathbb R} \times S^{d-1}$ that is branched over the cap of opening angle $2 \theta_0$.  Note the result \eqref{mainresult} and the steps leading up to it are essentially identical to a calculation and intermediate result derived in Ref.\ \cite{CardyHerzog} in 1+1 dimensions.  The difference is that in 1+1 dimensions, the two-point function $\langle \psi(z)\, \psi(z') \rangle_n$ can be evaluated for a general CFT through an appropriate conformal transformation, while in higher dimensions we only know how to evaluate $\langle \psi(z)\, \psi(z') \rangle_n$ in some special cases. 

In the case of free fields (and perhaps more generally) it makes sense to map this $n$-fold cover of the sphere to 
${\mathcal C}_n \times {\mathbb R}^{d-2}$ where ${\mathcal C}_n$ is a two dimensional cone of opening angle $2 \pi n$.    
In the case of a free theory, the two-point function 
$\langle \psi(y)\, \psi(y') \rangle_{1/m}$, where $y$, $y' \in {\mathcal C}_n \times {\mathbb R}^{d-2}$, 
can be evaluated by the method of images on a cone of opening angle $2\pi/m$ and then analytically continued to integer values of $1/m$.  Ref.\ \cite{Cardy} made successful use of this trick to calculate a limit of the mutual information for conformally coupled scalars.  We will use this same trick to look at thermal corrections to R\'enyi entropies for these scalars.  
%In $d=2$, the free fermion calculation agrees with earlier results  \cite{CardyHerzog,Azeyanagi:2007bj,Herzog:2013py}.
Taking the $n\to 1$ limit, we find complete agreement with entanglement entropy corrections computed in Ref.\ \cite{Herzog}.
(The method of images can also be used to study free fermions, but we leave such a calculation for future work.)
We verify the R\'enyi entropy corrections numerically by putting the system on a lattice. 
%\eqref{deltaSE} where $\Delta = (d-2)/2$, $g=1$, and $I_d$ is replaced by $I_{d-2}$.  
%(This last replacement is caused by boundary terms in the action for a conformally coupled scalar \cite{Herzog}.)

The paper is organized as follows. In Section \ref{sec:analytic}, we derive the result \eqref{mainresult} analytically. 
In Section \ref{sec:free}, we describe the conformal map to ${\mathcal C}_n \times \mathbb{R}^{d-2}$ and then work out the specific case of the conformally coupled scalar field.
Section \ref{sec:EE} computes thermal corrections to entanglement entropy by considering the $n \to 1$ limit of the R\'enyi entropy corrections of Section \ref{sec:free}.  The corrections agree with the results presented in Ref.\ \cite{Herzog}.
Section \ref{sec:numeric} provides a numerical check of the R\'enyi entropy corrections.
We conclude in Section \ref{sec:discussion} with a summary, discussion of related problems, and proposals for future research.
Appendix A provides details of a contour integral calculation of the scalar Green's function in $d=5$ dimensions, while Appendix B summarizes examples of thermal R\'enyi entropy corrections for the scalar for small values of $n$ and $d$.

\section{Analytical Calculation}
\label{sec:analytic}

We start with the thermal density matrix:
\be
  \rho = \frac{|0\rangle \langle 0| + \sum_i |\psi_i \rangle \langle \psi_i | \, e^{-\beta E_\psi} + \cdots}{1 + g\, e^{-\beta E_\psi} + \cdots}\, ,
\ee
where $|0\rangle$ stands for the ground state, while $|\psi_i\rangle\, (i = 1, \cdots, g)$  denote the first excited states. For a conformal field theory on $\mathbb{R} \times S^{d-1}$,
\be
  E_\psi = \frac{\Delta}{R}\, ,
\ee
where $\Delta$ is the scaling dimension of the operators that create the states $|\psi_i\rangle$, and $R$ is the radius of the sphere. From this expression one can calculate that
\begin{align}
  \textrm{tr}\, (\rho_A)^n & = \left(\frac{1}{1 + g\, e^{-\beta E_\psi} + \cdots} \right)^n \cdot \textrm{tr} \left[\textrm{tr}_B \left(|0\rangle \langle 0| + \sum_i |\psi_i\rangle \langle \psi_i| e^{-\beta E_\psi} + \cdots \right) \right]^n \nonumber\\
  {} & = \textrm{tr} \left(\textrm{tr}_B |0\rangle \langle 0| \right)^n \cdot \left[1 + \left(\frac{\textrm{tr} \left[\textrm{tr}_B \sum_i |\psi_i\rangle \langle \psi_i| \left(\textrm{tr}_B |0\rangle \langle 0| \right)^{n-1} \right]}{\textrm{tr} \left(\textrm{tr}_B |0\rangle \langle 0| \right)^n} - g \right) n\, e^{-\beta E_\psi} + \cdots \right]\, .
\end{align}
Then the thermal correction to the R\'enyi entropy is
\begin{align}\label{eq:corrRenyi}
  \delta S_n & \equiv S_n(T) - S_n(0)  \nonumber \\
    & = \frac{ n}{1-n} \sum_i \left(\frac{\textrm{tr} \left[\textrm{tr}_B  |\psi_i\rangle \langle \psi_i| \left(\textrm{tr}_B |0\rangle \langle 0| \right)^{n-1} \right]}{\textrm{tr} \left(\textrm{tr}_B |0\rangle \langle 0| \right)^n} - 1 \right) e^{-\beta E_\psi} + o \left(e^{- \beta E_\psi}\right)\, .
\end{align}
Hence, the crucial step is to evaluate the expression
\be\label{eq:2pt-fct}
  \frac{\textrm{tr} \left[\textrm{tr}_B |\psi_i\rangle \langle \psi_i| \left(\textrm{tr}_B |0\rangle \langle 0| \right)^{n-1} \right]}{\textrm{tr} \left(\textrm{tr}_B |0\rangle \langle 0| \right)^n} = 
\frac {\langle \psi_i(z) \psi_i(z') \rangle_n}{\langle \psi_i(z) \psi_i(z') \rangle_1}\, ,
\ee
which, using the operator-state correspondence, can be viewed as a two-point function on the $n$-fold covering of the space $\mathbb{R}\times S^{d-1}$.  (Let $z^\mu$ be our coordinate system on ${\mathbb R} \times S^{d-1}$.)  The $n$ copies are glued sequentially together along $A$.  Let $\tau$ be the time coordinate.  To create the excited state, we insert the operator $\psi_i$ in the far Euclidean past $\tau' = -i \infty$ of one of the copies of $\mathbb{R} \times S^{d-1}$.  Similarly, $\langle \psi_i |$ is created by inserting $\psi_i$ in the far future $\tau = i \infty$ of the same copy.
The two-point function $\langle \psi_i(z) \psi_i(z') \rangle_1$ is needed in the denominator in order to insure that $\langle \psi_i |$ has the correct normalization relative to $| \psi_i \rangle$.  

Our most general result is then
\be\label{eq:GenFormSn}
  \delta S_n = \frac{ n}{1-n}  \sum_i \left( \frac{\langle \psi_i(z)\, \psi_i(z') \rangle_n}{\langle \psi_i(z)\, \psi_i(z') \rangle_1} - 1 \right) e^{-\beta E_\psi} + o \left(e^{-\beta E_\psi}\right)\, .
\ee
Following from the analytic continuation formula \eqref{analyticcont}, 
the thermal correction to the entanglement entropy can be determined via
\be\label{eq:GenFormSE}
  \delta S_E = \lim_{n \to 1} \delta S_n\, .
\ee

\section{The free case}
\label{sec:free}

Evaluating the two-point function $\langle \psi_i(z) \psi_i(z') \rangle_n$ on an $n$-sheeted copy of
${\mathbb R} \times S^{d-1}$ is not simple for $n>1$.
Using a trick of Ref.\ \cite{Cardy},  we can evaluate $\langle \psi_i(z) \psi_i(z') \rangle_n$ for free CFTs.  The trick is to perform a conformal transformation that relates this two-point function to a two-point function on a certain conical space where the method of images can be employed.  As interactions spoil the linearity of the theory and hence the principle of superposition, we expect this method will fail for interacting CFTs.

It is convenient to break the conformal transformation into two pieces.
First, it is well known that ${\mathbb R} \times S^{d-1}$ is conformally related to Minkowski space 
(see the appendix of Ref.~\cite{Candelas:1978gf}):
\begin{eqnarray}
ds^2 &=& -dt^2 + dr^2 + r^2 d \Omega^2 \\
&=& \Omega^2 ( - d \tau^2 + d \theta^2 + \sin^2 \theta d \Omega^2) \ ,
\end{eqnarray}
where
\begin{eqnarray}
t \pm r &=& \tan \left( \frac{ \tau \pm \theta}{2} \right) \ , \\
\Omega &=& \frac{1}{2} \sec \left( \frac{ \tau + \theta}{2} \right) \sec \left( \frac{\tau - \theta}{2} \right) \ ,
\end{eqnarray}
and $d\Omega^2$ is a line element on a unit $S^{d-2}$ sphere.
Note that the surface $t=0$ gets mapped to $\tau = 0$, and on this surface $r = \tan( \theta/2)$.  Thus a cap on the sphere (at $\tau=0$) of opening angle $2\theta$ is transformed into a ball inside ${\mathbb R}^{d-1}$ (at $t=0$) of radius $r = \tan (\theta/2)$.  
%Our two point function has operators inserted in the far past and far future of Euclidean time in ${\mathbb R} \times S^d$.  
This coordinate transformation takes the operator insertion points $\tau = \pm i \infty$ in the far past and far future (with $\theta = 0$) to $t = \pm i$ (and $r=0$).

Then we should employ the special conformal transformation
\be
y^\mu = \frac{x^\mu - b^\mu x^2}{1 - 2 b \cdot x + b^2 x^2} \ ,
\ee
\be
ds^2 = dy^\mu dy^\nu \delta_{\mu\nu} = \frac{1}{(1 - 2 b \cdot x + b^2 x^2)^2} dx^\mu dx^\nu \delta_{\mu\nu} \ .
\ee
We let $x^0$ and $y^0$ correspond to Euclidean times.  We consider a sphere of radius $r$ in the remaining $d-1$ dimensions, centered about the origin.  If we set $b^1 = 1/r$ and the rest of the $b^\mu = 0$, this coordinate transformation will take a point on the sphere to infinity, specifically the point $x^\mu = (0,r,0, \ldots, 0)$.  The rest of the sphere will get mapped to a hyperplane with $y^1 = -r/2$.  We can think of the total geometry as a cone in the $(y^0, y^1)$ coordinates formed by gluing $n$-spaces together, successively, along the half plane $y^0 = 0$ and $y^1 < -r/2$.  
Let us introduce polar coordinates $(\rho, \phi)$ on the cone currently parametrized by $(y^0, y^1)$.  The tip of the cone $(y^0 , y^1) = (0,-r/2)$ will correspond to $\rho = 0$.
The insertion points $(\pm 1, 0, \ldots,0)$  for the operator $\psi_i$ get mapped to $(\pm 1, -1/r, 0,\ldots,0)/(1+1/r^2)$.  
  In polar coordinates, the insertion points of the $\psi_i$ are at $(r/2, \pm \theta)$.  
By a further rescaling and rotation, we can put the insertion points at $(1,2\theta,\vec 0)$ and $(1,0, \vec 0)$.

For primary fields $\psi_i(x)$, the effect of a conformal transformation on the ratio \eqref{eq:2pt-fct} is particulary simple.  
Let us focus on one of the $\psi_i = \psi$ and assume that it is a primary scalar field.
We have
\[
\psi(x) = \left(  \frac{1}{2} \sec \left( \frac{ \tau + \theta}{2} \right) \sec \left( \frac{\tau - \theta}{2} \right) \right)^{-\Delta} \psi(z) \ ,
\]
\[
\psi(y) = (1 - 2 b \cdot x + b^2 x^2)^{ \Delta} \psi(x) \ .
\]
We are interested in computing
\[
\frac {\langle \psi(z) \psi(z') \rangle_n}{\langle \psi(z) \psi(z') \rangle_1}\, ,
\]
where the subscript $n$ indicates this $n$-fold covering of the sphere, glued along the boundary of $A$.  In the ratio, the conformal factors relating the $z$ coordinates to the $x$ coordinates and the $x$ coordinates to the $y$ coordinates will drop out.  All we need pay attention to is where $y$ and $y'$ are in the cone of opening angle $2 \pi n$, which we have already done.  
For non-scalar and non-primary operators, the transformation rules are more involved.
%In our polar coordinate system 
%$(\rho, \theta, \vec r_\perp)$, we have that $y=(1,2\theta, \vec 0)$ and $y' = (1,0, \vec 0)$.

For free CFTs, $\langle \psi(y)\, \psi(y') \rangle_{1/m}$ can be evaluated for $m = 1, 2, 3, \ldots$ by the method of images.  For $n=1/m$, the conical space has opening angle $2\pi / m$. 
Let us assume we know the two-point function on ${\mathbb R}^{d}$: $\langle \psi(y_1) \psi(y_2) \rangle_1 = f(y_{12}^2)$.
Using the parametrization $y = (\rho, \theta, \vec r)$, the square of the distance between the points is
\be
y_{12}^2 = \rho_1^2 + \rho_2^2 - 2 \rho_1 \rho_2 \cos(\theta_{12}) + (\vec r_{12})^2 \ .
\ee
By the method of images, 
\be
\langle \psi(y_1) \psi(y_2) \rangle_{1/m} = \sum_{k=0}^{m-1} f \left(   \rho_1^2 + \rho_2^2 - 2 \rho_1 \rho_2 \cos(\theta_{12} + 2 \pi k / m) + (\vec r_{12})^2  \right) \ .
\ee
We are interested in two particular insertion points $y = (1,2\theta,\vec 0)$ and $y' = (1,0, \vec 0)$, for which the two point function reduces to
\be
\langle \psi(y) \psi(y') \rangle_{1/m} = \sum_{k=0}^{m-1} f \left(  2  -  2\cos(2 \theta+ 2 \pi k / m)   \right) \ .
\ee
Once we have obtained an analytic expression for all $m$, we can then evaluate it for integer $n=1/m$.

\subsection{The free scalar}
\label{sec:freescalar}

We now specialize to the case of a free scalar, for which the scaling form of the Green's function in flat Euclidean space
is $f(y^2) = y^{2-d}$. 
Our strategy will be to take advantage of recurrence relations that relate the Green's function in $d$ dimensions to $d+2$ dimensions.
Let us define 
\be
\label{defG}
G^{B}_{(n,d)}(2\theta) \equiv \langle \psi(y) \psi(y') \rangle_n \ .
\ee
%(The symbol $G^{F}_{(n,d)}$ is reserved for the free fermions we study in the next section.)
We need to compute the sum
\be
\label{eq:2ptFct-1}
  G^{B}_{(1/m,d)} (2\theta) = \langle \psi(y)\, \psi(y') \rangle_{1/m} = \sum_{k=0}^{m-1} \frac{1}{\left[2 - 2\, \textrm{cos}\left(2\theta + \frac{2 \pi k}{m} \right) \right]^{\frac{d-2}{2}}}\, .
\ee
As can be straightforwardly checked, this sum obeys the recurrence relation
\be\label{eq:recursion}
  G^B_{(1/m,d+2)}(\theta) = \frac{1}{(d-2) (d-1)} \left[\left(\frac{d-2}{2} \right)^2 + \frac{\partial^2}{\partial \theta^2} \right] G^B_{(1/m,d)} (\theta) \, .
\ee
The most efficient computation strategy we found is to compute $G^B_{(n,d)}$ for $d=3$ and $d=4$ and then to use the recurrence relation to compute the two point function in $d>4$.  (In $d=2$, the scalar is not gapped and there will be additional entanglement entropy associated with the degenerate ground state.)

To compute $G^{B}_{(n,4)}$, and more generally $G^B_{(n,d)}$ when $d$ is even, we introduce the generalized sum
\begin{align}\label{eq:generalFctS}
  f_a (m, \theta, z, \bar{z}) & \equiv \sum_{k=0}^{m-1} \frac{1}{|z - e^{i (\theta + 2 \pi k / m)} |^{2a}}\, ,
\end{align}
%As noticed in Ref.~\cite{Herzog}, there is a recurrence relation between the sums $\eqref{eq:2ptFct-1}$ in different dimensions. Let us define
%\begin{align}\label{eq:generalFctS}
%  f_a (m, \theta, z, \bar{z}) & \equiv \sum_{k=0}^{m-1} \frac{1}{|z - e^{i (\theta + 2 \pi k / m)} |^{2 a}}\, ,
%\end{align}
%such that
With this definition, we have the restriction that 
\be
\label{limit}
  \lim_{z, \bar{z} \to 1} f_{(d-2)/2} (m, \theta, z, \bar{z}) = 
  G^{B}_{(1/m,d)} (\theta)\ .
 % S_\alpha (m, \theta)\, ,
\ee
and the recurrence relation 
\be\label{eq:recurrence}
  \frac{\partial^2 f_a}{\partial z\, \partial \bar{z}} = a^2 \, f_{a + 1} (m, \theta, z, \bar{z})\, .
\ee
%Given $f_{1/2}(m, \theta, z, \bar z)$ and $f_1(m, \theta, z, \bar z)$, which correspond to $d=3$ and 4 respectively, 
%we can use the recurrence relation \eqref{eq:recurrence} to determine the Green's functions in all higher $d$.
%(As the scalar in $d=2$ is not gapped, the first interesting example is $d=3$.)

In the case $d=4$, we find that
\be
  f_1 (m, \theta, z, \bar{z}) 
   =
  \frac{m}{|z|^2 - 1} \left[\frac{1}{1 - z^{-m}\, e^{i m \theta}} + \frac{1}{1 - \bar{z}^{-m}\, e^{-i m \theta}} - 1 \right]\, .\label{eq:f1}
\ee
The two-point function can be obtained from Eq.~\eqref{eq:f1} by taking the limit $z, \bar{z} \to 1$:
\be
  G^{B}_{(n,4)}\, (\theta) = \lim_{z, \bar{z} \to 1} f_1 \left(\frac{1}{n}, \theta, z, \bar{z} \right) = \frac{1}{n^2 \left[2 - 2\, \textrm{cos} \left(\frac{\theta}{n} \right) \right]}\, .\label{eq:3+1DGreenFct}
\ee

For $d = 6$ dimensions the two-point function can be obtained by taking the $z, \bar{z} \to 1$ of  $f_2 (m, \theta, z, \bar{z})$:
\be
  G^{B}_{(n,6)}\, (\theta) =  \lim_{z, \bar{z} \to 1} f_2 \left(\frac{1}{n}, \theta, z, \bar{z} \right)
=
\frac{1 + \frac{2}{n^2} + (\frac{1}{n^2} - 1) \, \textrm{cos} (\frac{\theta}{n})}{3 n^2 \left[2 - 2\, \textrm{cos} (\frac{\theta}{n}) \right]^2}\, .\label{eq:5+1DGreenFct}
\ee
Applying the recurrence relation \eqref{eq:recursion} to the four dimensional result \eqref{eq:3+1DGreenFct} yields the same answer.
It is straightforward to calculate the Green's function in even $d>6$.

For $d=3$, we do not have as elegant expression for general $n$.  
Through a contour integral argument we will now discuss, for $n=1$, 2, and 3 we obtain
\begin{eqnarray}
\label{GB13}
  G^{B}_{(1,3)}\, (\theta) & = & \frac{1}{2 \sin \frac{\theta}{2}}\, ,\\
\label{GB23}
  G^{B}_{(2,3)}\, (\theta) & = & \frac{1 - \frac{\theta}{2\pi}}{2 \sin\frac{\theta}{2}}\, ,\\
\label{GB33} 
  G^{B}_{(3,3)}\, (\theta) & = & \frac{1}{2 \sin \frac{\theta}{2}} \left[ 1 - \frac{2}{\sqrt{3}} \sin \frac{\theta}{6} \right] \ . 
%  
 % \frac{\sqrt{3} \, \textrm{cos} \left(\frac{\pi}{6} \left(\frac{\theta}{2\pi} - 1 \right) \right) - 
 %3\, \textrm{sin} \left(\frac{\pi}{6} \left(\frac{\theta}{2\pi} - 1 \right) \right)}{6 \left[\textrm{cos} \left(\frac{\pi}{6} \left(\frac{\theta}{2\pi} - 
 %1 \right) \right) + \textrm{cos} \left(\frac{\pi}{2} \left(\frac{\theta}{2\pi} - 1 \right) \right) + 
 %\textrm{cos} \left(\frac{5 \pi}{6} \left(\frac{\theta}{2\pi} - 1 \right) \right) \right]}\, .
\end{eqnarray}
More general expressions for $G_{(n,d)}^B(\theta)$ with $d$ odd can be found in the next section.
Tables of thermal R\'enyi entropy corrections $\delta S_n$ for some small $d$ and $n$ are in Appendix B.

%Indeed, the sum \eqref{sjm}, while often divergent, also appears to have a more general validity. [[ well, does it or doesn't it? ]]

\subsection{Odd dimension and contour integrals}
\label{sec:odd}

Following Ref.~\cite{Cardy}, for $d$ an odd integer we express the Green's function in terms of an integral and evaluate it using the Cauchy residue theorem:
\begin{eqnarray}
  G_{(1/m,d)}^{B}(\theta) 
  %& = \sum_{k=0}^{m-1} \frac{1}{\left[2 - 2\, \textrm{cos} \left(\theta + \frac{2 \pi k}{m} \right) \right]^{\frac{d-1}{2}}} \nonumber\\
  {} & = &  \sum_{k=0}^{m-1} \frac{1}{\left[2\, \textrm{sin} \left(\frac{\theta}{2} + \frac{\pi k}{m} \right) \right]^{d-2}} \nonumber\\
  {} & = & \frac{1}{(2 \pi)^{d-2}}  \sum_{k=0}^{m-1}  \left[ \int_0^\infty dx \, \frac{x^{\frac{\theta}{2 \pi} + \frac{k}{m} - 1} }{1+x} \right]^{d-2} \nonumber\\
  {} & = & \frac{1}{(2 \pi)^{d-2}} \int_0^\infty dx_1 \cdots \int_0^\infty dx_{d-2} \left( \prod_{i=1}^{d-2} \frac{(x_i)^{\frac{\theta}{2\pi} - 1}}{1+x_i} \right) \left[\sum_{k=0}^{m-1} \left(\prod_{i=1}^{d-2} x_i \right)^{\frac{k}{m}} \right] \nonumber\\
  {} & = & \frac{1}{(2 \pi)^{d-2}} \int_0^\infty dx_1 \cdots \int_0^\infty dx_{d-2} \left( \prod_{i=1}^{d-2} \frac{(x_i)^{\frac{\theta}{2\pi} - 1}}{1+x_i} \right) \left[\frac{1 - \prod_{i=1}^{d-2} x_i}{1 - \left(\prod_{i=1}^{d-2} x_i \right)^{\frac{1}{m}}} \right] \, .\label{mycontourintegral}
 \end{eqnarray}
Then $G_{(n,d)}^B (\theta)$ is obtained by replacing $m$ with $\frac{1}{n}$.
%
%\be
 % G^{(n)}_{(d)} (1, \theta, 0) = \frac{1}{(2 \pi)^{d-1}} \int_0^\infty dx_1 \cdots \int_0^\infty dx_{d-1} \left( \prod_{i=1}^{d-1} 
 %\frac{(x_i)^{\frac{\theta}{2\pi} - 1}}{1+x_i} \right) \left[\frac{1 - \prod_{i=1}^{d-1} x_i}{1 - \left(\prod_{i=1}^{d-1} x_i \right)^n} \right] \, .
%\ee
While this integral expression is valid for all integers $d$, even and odd, for the even integers $d$ it is easier to evaluate the limit \eqref{limit} or use the recurrence relation \eqref{eq:recursion} with the (3+1) dimensional result \eqref{eq:3+1DGreenFct}. 

Using the integral \eqref{mycontourintegral}, the two-point function in $d=3$ becomes
\be
\label{Gn3contour}
  G^{B}_{(n,3)}\, ( \theta) = \frac{1}{2\pi} \int_0^\infty \frac{x^{\frac{\theta}{2\pi} - 1}\, (1-x)}{(1+x)\, (1-x^n)} dx\, ,\label{eq:2+1DGreenFct}
\ee
This integral can be done analytically. Essentially it is a contour integral with a branch point at $z = 0$ and some poles on the unit circle. For convenience, we can choose a branch cut to be the positive real axis, and a contour shown in Fig.~1.
    \begin{figure}[!htb]
      \begin{center}
        \includegraphics[width=0.6\textwidth]{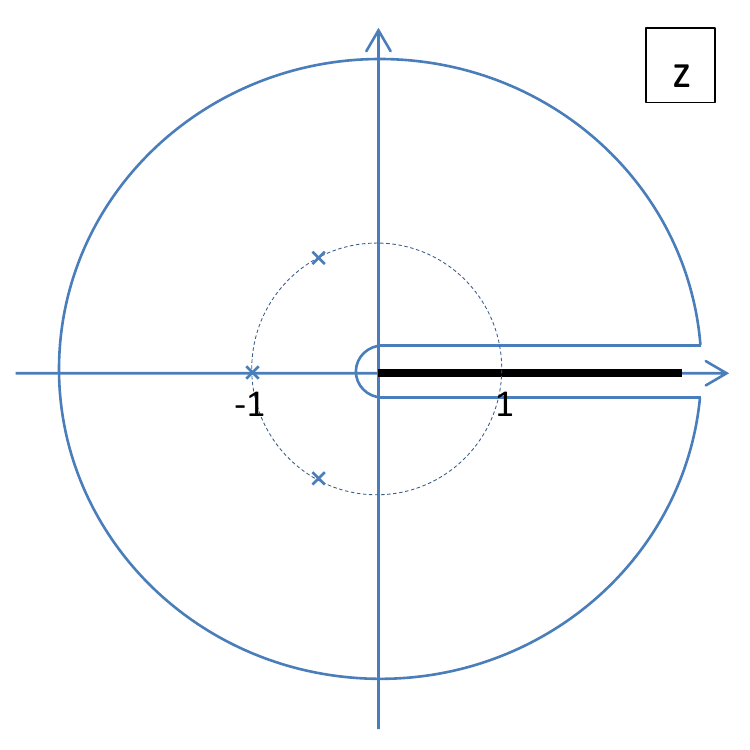}
      \end{center}
      \caption{\small The contour for $d=3$ dimensions and $n=3$}
    \end{figure}
For an even integer $n$, the poles are
\begin{displaymath}
  z = -1 \textrm{ is a double pole,}\quad z = e^{2 \pi i \frac{\ell}{n}}\,\, (\ell = 1, \cdots, \frac{n}{2} - 1, \frac{n}{2} + 1, \cdots, n - 1) \textrm{ are simple poles.}
\end{displaymath}
For an odd integer $n$, the poles are
\begin{displaymath}
  z = -1 \textrm{ and } z = e^{2 \pi i \frac{\ell}{n}}\,\, (\ell = 1, \cdots, n - 1) \textrm{ are all simple poles.}
\end{displaymath}
We emphasize that $z = 1$ is not a pole. Then for an even integer $n$:
\begin{align}
  G^B_{(n,d)} & = \frac{1}{2\pi} \int_0^\infty \frac{x^{\frac{\theta}{2\pi} - 1} (1 - x) }{(1 + x) (1 - x^n)} dx = \frac{i}{1 - e^{2 \pi i (\frac{\theta}{2 \pi} - 1)}} \, \sum_{\textrm{Poles}} \operatorname{Res} \\
  {} & = \frac{i}{1 - e^{2 \pi i (\frac{\theta}{2 \pi} - 1)}} \, \left[ - (-1)^{\frac{\theta}{2\pi} - 1} \left(\frac{\theta}{n\pi} - 1 \right) + \sum_{\ell=1,\, \ell \neq \frac{n}{2}}^{n-1} \frac{e^{2 \pi i \frac{\ell}{n} (\frac{\theta}{2\pi} - 1)}}{1 + e^{2 \pi i \frac{\ell}{n}}} \left(\prod_{j=1,\, j \neq \ell}^{n-1} \frac{1}{e^{2 \pi i \frac{\ell}{n}} - e^{2 \pi i \frac{j}{n}}} \right) \right]\, \nonumber \\
{}
  & = \frac{1}{2 \sin \frac{\theta}{2}} \left[ 1 - \frac{\theta}{\pi n} - \frac{i}{n} \sum_{\ell=1, \ell \neq n/2}^{n-1} e^{i \theta(\frac{\ell}{n} - \frac{1}{2} )} \tan \frac{\pi \ell}{n} \right]
  , \nonumber
\end{align}
while for an odd integer $n$:
\begin{align}
  G^B_{(n,d)} &= \frac{1}{2\pi} \int_0^\infty \frac{x^{\frac{\theta}{2\pi} - 1} (1 - x) }{(1 + x) (1 - x^n)} dx = \frac{i}{1 - e^{2 \pi i (\frac{\theta}{2 \pi} - 1)}} \, \sum_{\textrm{Poles}} \operatorname{Res} \\
  {} & = \frac{i}{1 - e^{2 \pi i (\frac{\theta}{2 \pi} - 1)}} \, \left[(-1)^{\frac{\theta}{2 \pi} - 1} + \sum_{\ell=1}^{n-1} \frac{e^{2 \pi i \frac{\ell}{n} (\frac{\theta}{2\pi} - 1)}}{1 + e^{2 \pi i \frac{\ell}{n}}} \left(\prod_{j=1,\, j \neq \ell}^{n-1} \frac{1}{e^{2 \pi i \frac{\ell}{n}} - e^{2 \pi i \frac{j}{n}}} \right) \right]\, \nonumber \\
 {}
  & =
   \frac{1}{2 \sin \frac{\theta}{2}} \left[ 1 - \frac{i}{n} \sum_{\ell=1}^{n-1} e^{i \theta(\frac{\ell}{n} - \frac{1}{2} )} \tan \frac{\pi \ell}{n} \right]
   . \nonumber
\end{align}
Therefore, for $d = 3$ dimensions the results for $n = 1, 2, 3$ are Eqs.~\eqref{GB13}--\eqref{GB33}.

Given the results for  $d=3$ dimensions, the two-point functions for $d=5$ dimensions can be obtained by using the recurrence relation \eqref{eq:recursion}:
\begin{eqnarray}
\label{GB15}
  G^{B}_{(1,5)}\, (\theta) & = & \frac{1}{\left( 2  \sin \frac{\theta}{2} \right)^3}
  %- (n-1) \frac{3\pi}{128} \frac{1}{\cos^4 \frac{\theta}{4}} + \mathcal{O}(n-1)^2
  \, ,\\
  \label{GB25}
%  G^{B}_{(2,5)}\, (\theta) & = & \frac{2 \pi (\frac{\theta}{2\pi} - 1) - 
%\textrm{sin} \left(2 \pi (\frac{\theta}{2\pi} - 1) \right)}{16\, \pi\, \textrm{sin}^3 \left(\pi (\frac{\theta}{2\pi} - 1) \right)}\, ,\\
 G^B_{(2,5)}\, (\theta) &=& \frac{2 \pi - \theta + \sin \theta}{2 \pi \left( 2\sin  \frac{\theta}{2} \right)^3} \ , \\
  \label{GB35}
G^B_{(3,5)}\, (\theta) &=& 
\frac{1}{108 \left( 2 \sin \frac{\theta}{2} \right)^3 } 
\left[ 
108 - 70 \sqrt{3} \sin \left( \frac{\theta}{6} \right) + 7 \sqrt{3} \sin \left( \frac{5 \theta}{6} \right) + 5 \sqrt{3} \sin \left( \frac{7 \theta}{6} \right) \right] \ .
%
%  G^{B}_{(3,5)}\, (\theta) & = & \frac{1}{864 \left[1 + 2\, \textrm{cos} \left(\frac{2}{3} \pi (\frac{\theta}{2 \pi} - 1) \right) \right]^3} \Bigg[88 \sqrt{3} \, \textrm{cos} \left(\frac{2}{3} \pi (\frac{\theta}{2 \pi} - 1) \right) + 20 \sqrt{3} \, \textrm{cos} \left(\frac{4}{3} \pi (\frac{\theta}{2 \pi} - 1) \right) \nonumber\\
%  {} & & \qquad \qquad \qquad \qquad \qquad \qquad \qquad - 96 \, \textrm{sin} \left(\frac{2}{3} \pi (\frac{\theta}{2 \pi} - 1) \right) - 60 \, \textrm{sin} \left(\frac{4}{3} \pi (\frac{\theta}{2 \pi} - 1) \right) \nonumber\\
%  {} & & \qquad \qquad \qquad \qquad \qquad \qquad \qquad - 27 \, \textrm{tan} \left(\frac{1}{6} \pi (\frac{\theta}{2 \pi} - 1) \right) \left(2 + \textrm{sec}^2 \left(\frac{1}{6} \pi (\frac{\theta}{2 \pi} - 1) \right) \right)
%  \nonumber \\
%  {} & &
%  \hskip 4in + 102 \sqrt{3} \Bigg]\, .
\end{eqnarray}
In Appendix A, we also compute the two-point function for $d=5$ dimensions and $n = 1, 2, 3$ by directly evaluating the contour integral
\be
  G^{B}_{(n,5)}\, ( \theta) = \frac{1}{(2\pi)^3} \int_0^\infty dx\, \int_0^\infty dy\, \int_0^\infty dz\, \frac{(xyz)^{\frac{\theta}{2\pi} - 1}\, (1 - xyz)}{(1+x)\, (1+y)\, (1+z)\, \left(1 - (xyz)^n \right)}\, ,\label{eq:4+1DGreenFct}
\ee
and the results are exactly the same.

\section{Thermal Corrections to Entanglement Entropy}
\label{sec:EE}

General results for thermal corrections to entanglement entropy were given in Ref.\ \cite{Herzog}.  Here we will verify these general results in arbitrary dimension for the specific case of a conformally coupled scalar.  To perform the check, we will use the fact that 
the $n\to 1$ limit of the R\'enyi entropies yields the entanglement entropy.  

The Green's function $G_{(n,d)}^{B}(\theta)$ has an expansion near $n=1$ of the form
\be
G_{(n,d)}^{B}( \theta) = G_{(1,d)}^{B}(\theta) + (n-1) \delta G^B_{(d)}( \theta) + \mathcal{O}(n-1)^2\ .
\ee
From the definition \eqref{defG} and the main result \eqref{eq:GenFormSn}, we have that
\be
\delta S_E = - \frac{\delta G^B_{(d)}(2 \theta)}{G_{(1,d)}^{B}(2\theta)} e^{-\beta E_\psi} + o(e^{-\beta E_\psi})\ .
\label{ourdSE}
\ee
Note that $\delta G^B_{(d)}(\theta)$ will also satisfy the recurrence relation \eqref{eq:recursion}.  Thus it is enough to figure out the thermal corrections for the smallest dimensions $d=3$ and $d=4$.  The result in $d>4$ will then follow from the recurrence.

Let us check that the expression \eqref{ourdSE} agrees with Ref.\ \cite{Herzog} in the cases $d=3$ and $d=4$. 
In the case $d=3$, 
we can evaluate the relevant contour integral \eqref{Gn3contour} in the limit $n \to 1$:
\begin{eqnarray}
 G^{B}_{(n,3)}\, (\theta) &=& 
 \frac{1}{2\pi} \int_0^\infty \frac{x^{\frac{\theta}{2\pi}-1}}{1+x} dx + \frac{n-1}{2\pi} \int_0^\infty \frac{x^{\frac{\theta}{2\pi}} \log x}{1-x^2}
 + \mathcal{O}(n-1)^2 
 \nonumber \\
 &=&
 \frac{1}{2} \frac{1}{\sin \frac{\theta}{2}} - (n-1) \frac{\pi}{8} \frac{1}{\cos^2 \frac{\theta}{4}} + \mathcal{O}(n-1)^2 \ .
\label{d2none}
\end{eqnarray}
From Eqs.~\eqref{ourdSE} and \eqref{d2none}, we then have 
\be
\label{deltaSEd3}
\delta S_E = \frac{\pi}{2} \tan \left( \frac{\theta}{2} \right)  e^{-\beta/2R} + o(e^{-\beta/2R}) \ .
\ee
For $d=4$, 
we expand Eq.~\eqref{eq:3+1DGreenFct} near $n=1$:
\be
G^{B}_{(n,4)}\, ( \theta) =  \frac{1}{4 \sin^2 \frac{\theta}{2}} \left( 1 + (n-1) \left(-2 + \theta \cot \frac{\theta}{2} \right) + \mathcal{O}(n-1)^2 \right) \ .
\label{d3none}
\ee
%The expansion of $G^B_{(n,d)} (\theta)$ satisfies the recurrence relation \eqref{eq:recursion}, 
%so we can derive the expansions for all even $d$'s near $n=1$ from Eq.~\eqref{d3none}.
We find from Eqs.~\eqref{ourdSE}  and \eqref{d3none} that 
\be
\label{deltaSEd4}
\delta S_E = 2 (1 - \theta \cot \theta) e^{-\beta/R} + o(e^{-\beta/R})\ .
\ee
The expressions \eqref{deltaSEd3} and \eqref{deltaSEd4} are precisely the results found for the conformally coupled scalar in Ref.\ \cite{Herzog} in $d=3$ and $d=4$ respectively.

Indeed, for general $d$, the result in Ref. \cite{Herzog} for the conformally coupled scalar is
\be
\delta S_E = \frac{d-2}{2} I_{d-2} (\theta) e^{-\beta (d-2)/2R} + o(e^{-\beta (d-2)/2R}) \ .
\ee
where the definition \eqref{Idintegral} of $I_d(\theta)$ was given in the introduction. 
%\be
%I_d(\theta) \equiv 2 \pi \frac{\Vol(S^{d-2})}{\Vol(S^{d-1})} \int_0^{\theta} \frac{ \cos x - \cos \theta}{\sin \theta} \sin^{d-2} x \, dx \ .
%\label{Idintegral}
%\ee
If our result \eqref{ourdSE} for the thermal correction is correct, we can relate $I_d(\theta)$ and $\delta G^B_{(d)}(\theta)$:
\be
\delta G^B_{(d)}(2 \theta) = - \frac{d-2}{2}  (2 \sin \theta)^{2-d} I_{d-2}(\theta) \ ,
\label{tentativeid}
\ee
where we have used the fact that $G^B_{(1,d)}(2 \theta) = (2 \sin \theta)^{2-d}$.

To check that our thermal corrections are correct for general $d$, we will use a roundabout method.  In Ref.\ \cite{Herzog}, it was also found that the function $I_d(\theta)$ satisfies a recurrence relation
\be
I_{d}(\theta) - I_{d-2}(\theta) = - 2\pi \frac{ \Vol(S^{d-2})}{\Vol(S^{d-1})} \frac{\sin^{d-2} \theta}{(d-1)(d-2)} \ .
\ee
We will use our recurrence relation \eqref{eq:recursion} and the tentative identification \eqref{tentativeid}
to replace $I_d(\theta)$ with $I_{d-2}(\theta)$ in the above expression:
\begin{eqnarray}
I_d(\theta) &=& - \frac{2}{d} (2 \sin \theta)^{d} \, \delta G^B_{(d+2)} (2 \theta) \nonumber \\
&=& - \frac{2  (2 \sin \theta)^{d} }{d(d-1)(d-2)} \left[ \left( \frac{d-2}{2} \right)^2 + \frac{1}{4} \frac{\partial^2}{\partial \theta^2} \right] \delta G^B_{(d)}(2 \theta) \nonumber \\
&=&\frac{  (2 \sin \theta)^{d} }{4d(d-1)} \left[ \left(d-2 \right)^2 + \frac{\partial^2}{\partial \theta^2} \right]  (2 \sin \theta)^{2-d}I_{d-2}(\theta) \ .
\end{eqnarray}
Then we have checked that the resulting differential equation in $I_{d-2}(\theta)$ is solved by the integral formula \eqref{Idintegral}.

\section{Numerical Check}
\label{sec:numeric}

%We will use a variant of the technique pioneered by Ref.\ \cite{Srednicki:1993im} 
%to compute the corrections to the R\'enyi entropies directly.  
We check numerically the thermal R\'enyi entropy corrections obtained in section \ref{sec:freescalar}.
The algorithm we use was described in detail in Ref.\ \cite{Herzog}, so we shall be brief. 
(The method is essentially that of
Ref.\ \cite{Srednicki:1993im}.)
The action for a conformally coupled scalar on ${\mathbb R} \times S^{d-1}$ is
\be
  S = -\frac{1}{2} \int d^{d} x\, \sqrt{-g} \left[(\partial_\mu \phi) (\partial^\mu \phi) + \xi \, \mathcal{R} \, \phi^2 \right]\, ,
\ee
where $\xi$ is the conformal coupling $ \xi = (d-2)/4(d-1)$ and ${\mathcal R}$ is the Ricci scalar curvature.
Given that the region $A$ can be characterized by the polar angle $\theta$ on $S^{d-1}$, we write the Hamiltonian as a sum $H = \sum_{\vec l} H_{\vec l}$,
 where we have replaced all the other angles on $S^{d-1}$ by corresponding angular momentum quantum numbers $|l_1| \leq l_2 \leq \cdots \leq l_{d-2}\equiv m$.  The individual Hamiltonians take the form
 \be
 H_{\vec l} = \frac{1}{2 R^2} \int_0^\pi \left\{ R^2 \Pi_{\vec l}^2 - \Phi_{\vec l} \partial_\theta^2 \Phi_{\vec l} + \frac{1}{4}(2m+d-2)(2m+d-4) \frac{\Phi_{\vec l}^2}{\sin^2 \theta} \right\} d \theta \ .
 \ee
 
 It is convenient to discretize $H_{\vec l}$.  In $d\geq 4$, we introduce a lattice in $\theta$, while in $d=3$, a lattice in $\cos \theta$ appears to work better.  The entanglement and R\'enyi entropies can then be expressed in terms of two-point functions restricted to the region $A$.
 In particular, the R\'enyi entropy can be expressed as
\be
  S_n (T) = S_n (0) + \sum_{m=1}^\infty \, \textrm{dim}(m)\, S_n^{(m)}\, ,\label{eq:ExpRE}
\ee
where
\[
\dim(m) = {d+m-2 \choose d-2} - {d+m-4 \choose d-2} \ ,
\]
and
\be
  S_n^{(m)} = \frac{1}{1-n} \textrm{log}\, \textrm{tr} \left[\left(C_m + \frac{1}{2} \right)^n - \left(C_m - \frac{1}{2} \right)^n \right]\, .
\ee
The matrix $C_m (\theta_1, \theta_2)$ has a continuum version
\be
  C_m (\theta_1, \theta_2)^2 = \int_0^{\theta_0} d\theta\, \langle \Phi_{\vec{l}} (\theta_1)\, \Phi_{\vec{l}} (\theta) \rangle \langle \Pi_{\vec{l}} (\theta)\, \Pi_{\vec{l}} (\theta_2) \rangle\, .
\ee
The thermal two-point functions have the following expressions:
\begin{eqnarray}
  \langle \Phi_{\vec{l}} (\theta)\, \Phi_{\vec{l}} (\theta') \rangle & = & \frac{1}{2} \sum_{l=m}^\infty U_l (\theta) \, \frac{1}{\omega_l} \, \textrm{coth}\frac{\omega_l}{2T}\, U_l (\theta')\, ,\label{eq:therm2pt-1}\\
  \langle \Pi_{\vec{l}} (\theta)\, \Pi_{\vec{l}} (\theta') \rangle & = & \frac{1}{2} \sum_{l=m}^\infty U_l (\theta) \, \omega_l \, \textrm{coth}\frac{\omega_l}{2T}\, U_l (\theta')\, ,\label{eq:therm2pt-2}
\end{eqnarray}
where
$\omega_l \equiv \frac{1}{R} \left(l + \frac{d-2}{2} \right)$.
In the continuum limit, the matrix $U_{l} (\theta)$ is an orthogonal transformation involving associated Legendre functions whose explicit form is given in Ref.~\cite{Herzog}.  In practice, we use the discretized version of $U_l(\theta)$ that follows from the discretized $H_{\vec l}$.

As discussed in Ref.~\cite{CardyHerzog, Herzog}, if the limit $\theta_0 \to \pi$ is taken first, the leading correction to $\delta S_n$ comes from the thermal R\'enyi entropy instead of from the entanglement:
\be
  \delta S_n = \left[-g \frac{n}{1-n} + \mathcal{O} \left(1 - \frac{\theta_0}{\pi} \right)^{2 \Delta} \right] e^{-\Delta / RT} + o \left(e^{-\Delta / RT} \right)\, .
\ee
Indeed, when $\pi - \theta$ is small compared to $RT$, the R\'enyi entropy looks like the thermal R\'enyi entropy and approaches it in the limit $\theta \to \pi$. To isolate the $e^{-\Delta / RT}$ dependence of $\delta S_n$ analytically, we can expand the $\textrm{coth}$-function in the thermal two-point functions \eqref{eq:therm2pt-1} and \eqref{eq:therm2pt-2}. In principle, one can evaluate Eq.~\eqref{eq:ExpRE} to obtain $\delta S_n (T)$. Since we are interested in the low temperature limit, the contributions from $S_n^{(m)}\,\, (m>0)$ to $\delta S_n (T)$ are exponentially suppressed compared with $S_n^{(0)}$. Therefore, in the limit of small $T$, we obtain the expansion of Eq.~\eqref{eq:ExpRE}:
%[[ changed prefactor ]]
\be
  \delta S_n = \frac{n}{2(n-1)} \textrm{tr} \left[\delta C_0 \cdot C_0^{-1} \cdot \frac{(C_0 + \frac{1}{2})^{n-1} - (C_0 - \frac{1}{2})^{n-1}}{(C_0 + \frac{1}{2})^n - (C_0 - \frac{1}{2})^n} \right] e^{-\omega_0 / T} + \cdots\, ,
\ee
where
\begin{align}
  \delta C_m (\theta_1, \theta_2) & \equiv \int_0^{\theta_0} d\theta\, \left[\langle \Phi_{\vec{l}} (\theta_1)\, \Phi_{\vec{l}} (\theta) \rangle \, \delta \Pi_m (\theta, \theta_2) + \delta \Phi_m (\theta_1, \theta)\, \langle \Pi_{\vec{l}} (\theta)\, \Pi_{\vec{l}} (\theta_2) \rangle \right]\, ,\\
  \delta \Phi_m (\theta, \theta') & \equiv U_m (\theta)\, \frac{1}{\omega_m}\, U_m (\theta')\, ,\\
  \delta \Pi_m (\theta, \theta') & \equiv U_m (\theta)\, \omega_m\, U_m (\theta')\, .
\end{align}
 Some results of $\delta S_n$ in different dimensions are shown in Figs.~2 -- 5. 
 To diagonalize the matrices with enough accuracy, high precision arithmetic is required.
 %As discussed in Ref.~\cite{Herzog}, for $(d+1)=(2+1)$ dimensions a grid in $u = \textrm{cos} \theta$ 
 %works much better, so we plot $\Delta S_n$ vs. $\textrm{cos} \theta$ in Fig.~2.

\begin{figure}[!htb]
\centering
\begin{minipage}{.5\textwidth}
  \centering
  \includegraphics[width=0.86\linewidth]{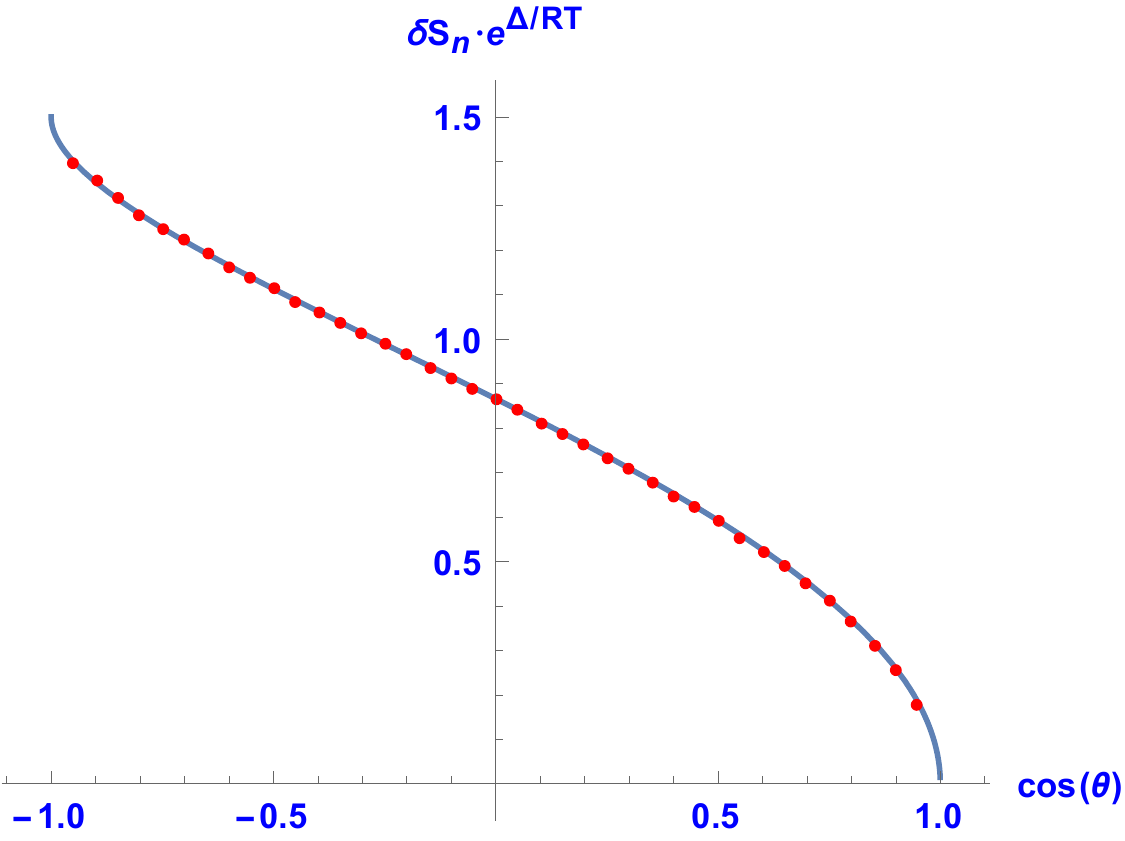}
  \caption{\small $\delta S_{n=3}$ in $(2+1)$ D, $400$ grid points}
\end{minipage}%
\begin{minipage}{.5\textwidth}
  \centering
  \includegraphics[width=0.99\linewidth]{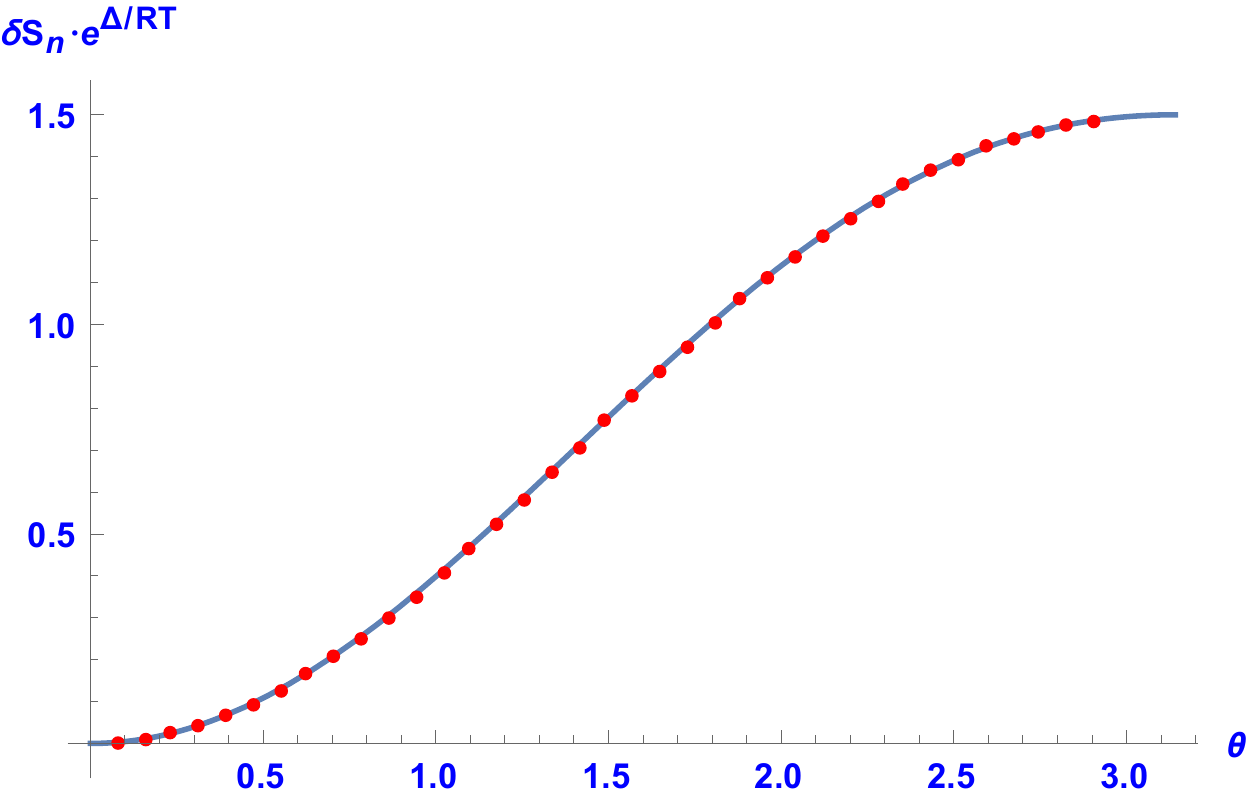}
  \caption{\small $\delta S_{n=3}$ in $(3+1)$ D, $400$ grid points}
\end{minipage}
\end{figure}

\begin{figure}[!htb]
\centering
\begin{minipage}{.5\textwidth}
  \centering
  \includegraphics[width=0.94\linewidth]{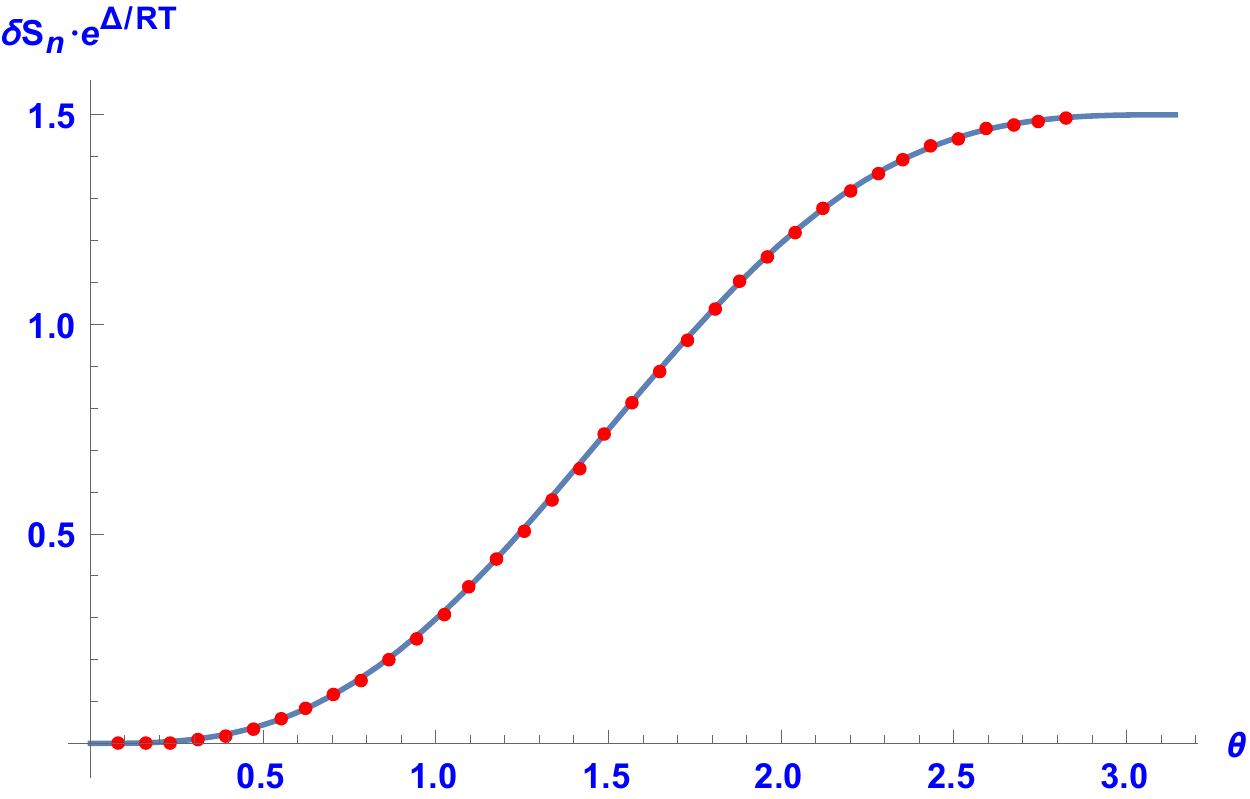}
  \caption{\small $\delta S_{n=3}$ in $(4+1)$ D, $400$ grid points}
\end{minipage}%
\begin{minipage}{.5\textwidth}
  \centering
  \includegraphics[width=0.95\linewidth]{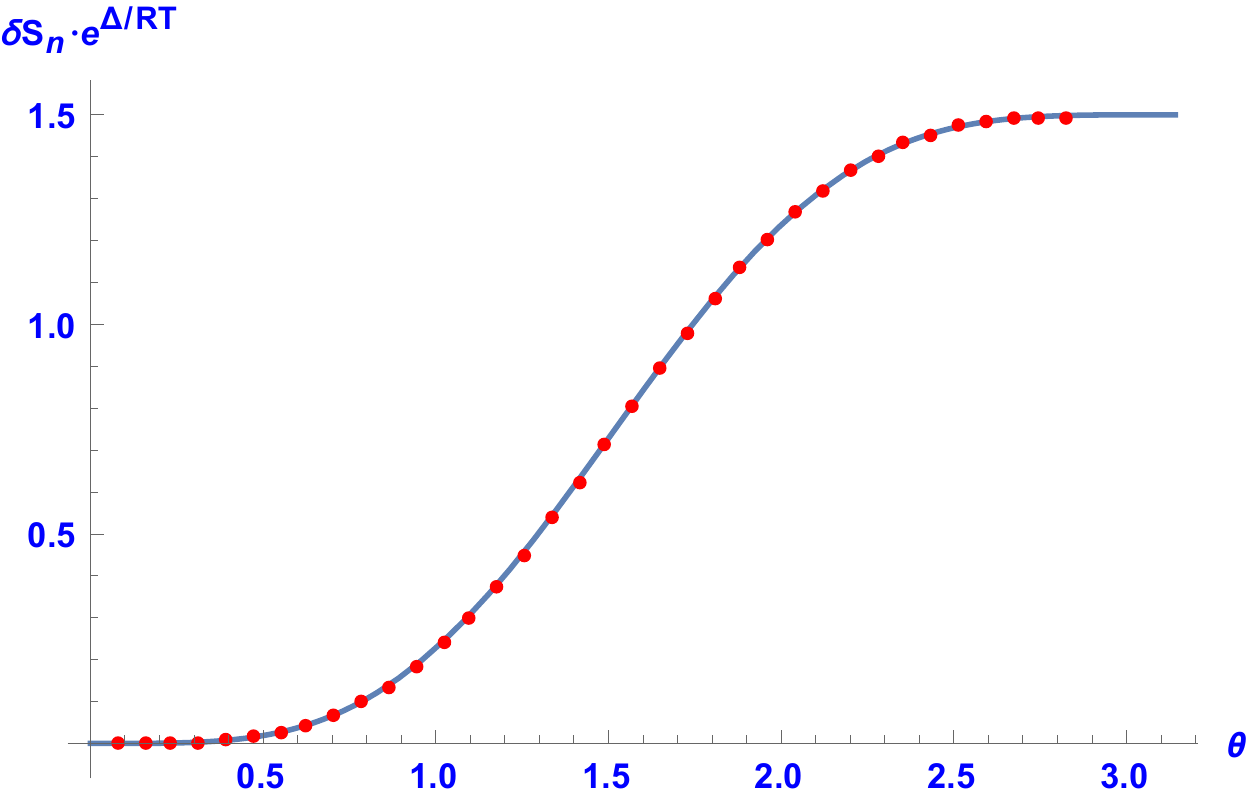}
  \caption{\small $\delta S_{n=3}$ in $(5+1)$ D, $400$ grid points}
\end{minipage}
\end{figure}

\section{Discussion}
\label{sec:discussion}

Our main result provides a way to calculate the leading thermal correction to a specific kind of R\'enyi entropy for a CFT.  In particular, the CFT should live on ${\mathbb R} \times S^{d-1}$, and the region is a cap on the sphere with opening angle $2 \theta$.  
We demonstrated that this correction is equivalent to knowing the two-point function on a certain conical space of the operator that creates the first excited state.  
In the case of a conformally coupled free scalar, the scalar field itself creates the first excited state, and the two point function can be computed by the method of images.  In the $n \to 1$ limit, R\'enyi entropy becomes entanglement entropy, and we were able to show that our results agree with Ref.~\cite{Herzog}.  We were also able to check our thermal corrections for $n>1$ numerically, using a method based on Ref.~\cite{Srednicki:1993im}.  

We would like to make two observations about our results.  The first is that our thermal R\'enyi entropy corrections are often but not always invariant under the replacement $\theta \to 2 \pi n - \theta$.   (The exceptions are $\delta S_n$ for even $n$ and odd $d$.)  A similar observation was made in Ref.~\cite{CardyHerzog} in the 1+1 dimensional case.  There, the invariance could be explained by moving twist operators around the torus (or cylinder).   The branch cut joining two twist operators is the same cut along which the different copies of the torus are glued together.  By moving a twist operator $n$ times around the torus, $n$ branch cuts are equivalent to nothing while $n-1$ branch cuts are equivalent to a single branch cut that moves one down a sheet rather than up a sheet.
Perhaps in higher dimensions 
the invariance can be explained in terms of surface operators that glue the $n$ copies of $S^1 \times S^{d-1}$ together. It is not clear to us how to generalize the argument.   It is tempting to speculate that the invariance is spoiled in odd dimensions (even dimensional spheres) because only odd dimensional spheres are Hopf fibrations over projective space.

The second observation is that the leading corrections to $\delta S_n$ for small caps $\theta \ll 1$ have a power series expansion that starts with the terms $a \theta^{d-2} + b \theta^{d} + \ldots$.  In 1+1 dimensions, the power series starts with $\theta^2$ \cite{CardyHerzog}.  When we bring two twist operators together, the twist operators can be replaced by their operator product expansion, a leading term of which is the stress tensor.  The $\theta^2$ term in $\delta S_n$ comes from a three point function of the stress tensor with the operators that create and annihilate the first excited state.  The two in the exponent of $\theta^2$ comes from the scaling dimension of the stress tensor, and the coefficient of the $\theta^2$ can be related to the scaling dimension of the twist operators \cite{CardyHerzog}. 
In our higher dimensional case, we can replace the surface operator along the boundary of the cap by an operator product expansion at a point.  Because of Wick's theorem, the leading operator that can contribute to $\delta S_n$ will be $\phi^2$ which has dimension $d-2$.  The subleading $\theta^d$ term may come from the stress tensor and descendants of $\phi^2$.  A more detailed analysis might shed some light on the structure of these surface operators.\footnote{%
 See Refs.~\cite{Cardy, Shiba:2012np, Hung,Casini:2008wt} for related work on higher dimensional analogs of twist operators.
}

 In addition to developing the above observations, we give a couple of projects for future research.
One would be to compute these thermal corrections for free fermions.  The two point function on this conical space can quite likely be computed.  It would be interesting to see how the results compare to the scalar.  Given the importance of boundary terms for the scalar, it would also be nice to get further confirmation of the general story for thermal corrections to entanglement entropy presented in Ref.\ \cite{Herzog}.

Another interesting project would be to see how to obtain these results holographically.
As the corrections are subleading in a large central charge (or equivalently large $N$) expansion, they would not be captured by the Ryu-Takayanagi formula \cite{Ryu}.
However, it may be possible to generalize the computation in $d=2$ \cite{Barrella:2013wja} to $d>2$.  
Finally, it would be interesting to see what can be said about negativity in higher dimensions.  See Ref.\ \cite{Calabrese:2014yza} for the two dimensional case.

\vskip 0.1in

\section*{Acknowledgments}
We would like to thank Michael Spillane, Pin-Ju Tien, and Ricardo Vaz for useful discussions. 
C.~H. and J.~N. were supported in part by the National Science Foundation under Grant No.\ PHY13-16617.  C.~H. thanks the Sloan Foundation for partial support.

\appendix

\section{Two-Point Functions in $d=(4+1)$ Dimensions}
In this appendix, we compute the two-point function for $d=(4+1)$ dimensions given by Eq.~\eqref{eq:4+1DGreenFct}. In contrast to Eq.~\eqref{eq:2+1DGreenFct}, Eq.~\eqref{eq:4+1DGreenFct} is a multi-variable contour integral and we need to do some changes of variables first. The procedure used here can be applied in higher dimensions and for any integer $n\geq 1$.
In $d=5$, we find
\begin{eqnarray}
  G^{B}_{(n,5)}\, (\theta) & = & \frac{1}{(2\pi)^3} \int_0^\infty dx\, \int_0^\infty dy\, \int_0^\infty dz\, \frac{(xyz)^{\frac{\theta}{2\pi} - 1}\, (1 - xyz)}{(1+x)\, (1+y)\, (1+z)\, \left(1 - (xyz)^n \right)} \nonumber\\
  {} & = & \frac{1}{(2\pi)^3} \int_0^\infty dx\, \int_0^\infty dy\, \int_0^\infty \frac{1}{xy} dz'\, \frac{z'^{\frac{\theta}{2\pi} - 1} (1-z')}{(1+x) (1+y) (1+\frac{z'}{xy}) (1-z'^n)} \nonumber\\
  {} & = & \frac{1}{(2\pi)^3} \int_0^\infty dx\, \int_0^\infty dy\, \int_0^\infty dz'\, \frac{z'^{\frac{\theta}{2\pi} - 1} (1-z')}{(1+x) (1+y) (xy+z') (1-z'^n)} \nonumber\\
  {} & = & \frac{1}{(2\pi)^3} \int_0^\infty dx\, \int_0^\infty \frac{1}{x} dy'\, \int_0^\infty dz'\, \frac{z'^{\frac{\theta}{2\pi} - 1} (1-z')}{(1+x) (1+\frac{y'}{x}) (y'+z') (1-z'^n)} \nonumber\\
  {} & = & \frac{1}{(2\pi)^3} \int_0^\infty dx\, \int_0^\infty dy'\, \int_0^\infty dz'\, \frac{z'^{\frac{\theta}{2\pi} - 1} (1-z')}{(1+x) (x+y') (y'+z') (1-z'^n)} \nonumber\\
  {} & = & \frac{1}{(2\pi)^3} \int_0^\infty dx\, \int_0^\infty dy\, \int_0^\infty dz\, \frac{z^{\frac{\theta}{2\pi} - 1} (1-z)}{(1+x) (x+y) (y+z) (1-z^n)}\, ,\label{eq:intermediateInt}
\end{eqnarray}
where
\be
  z' \equiv xyz\, ,\quad y' \equiv xy\, ,
\ee
and we drop the ${}'$ in the last line. Performing the integration over $x$ and $y$ in Eq.~\eqref{eq:intermediateInt}, we obtain
\be
  G_{(n,5)}^{B}\, (\theta) = \frac{1}{2 (2 \pi)^3} \int_0^\infty dz\, \frac{z^{} \left(\pi^2 + (\textrm{log} z)^2 \right) (z - 1)}{(1 + z) (z^n - 1)}\, .
\ee
This integral can be done analytically by choosing the same branch cut and contour used in the $d=(2+1)$ dimensional case discussed in Section \ref{sec:odd}; the poles are exactly the same. The result for $n=1$ is Eq.\ \eqref{GB15}.
%\be
%  G^{B}_{(1,5)}\, (\theta) = \frac{1}{8 \, \textrm{sin}^3 \left(\frac{\theta}{2} \right)}\, .
%\ee
For $n=2$ the result is Eq.~\eqref{GB25}.
%\be
%  G^{B}_{(2,5)}\, (\theta) = \frac{2 \pi (\frac{\theta}{2\pi} - 1) - \textrm{sin} \left(2 \pi (\frac{\theta}{2\pi} - 1) \right)}{16\, \pi\, \textrm{sin}^3 \left(\pi (\frac{\theta}{2\pi} - 1) \right)}\, .
%\ee
To obtain this result, one needs the following intermediate results
\begin{align}
  \int_0^\infty dz\, \frac{z (z-1)}{(1+z) (z^2 - 1)} & = \frac{\pi (\frac{\theta}{2\pi} - 1)}{\textrm{sin} \left[\pi \left(\frac{\theta}{2 \pi} - 1 \right) \right]}\, ,\\
  \int_0^\infty dz\, \frac{z \, \textrm{log}\, z\, (z-1)}{(1+z) (z^2-1)} & = \frac{\pi \left[1 - \pi (\frac{\theta}{2 \pi} - 1)\, \textrm{cot} \left(\pi (\frac{\theta}{2\pi} - 1) \right) \right]}{\textrm{sin} \left(\pi (\frac{\theta}{2\pi} - 1) \right)}\, .
\end{align}
Similarly, for $n=3$ one can follow exactly the same procedure and find Eq.~\eqref{GB35}.
%\begin{align}
%  G^{B}_{(3,5)}\, (\theta) & = \frac{1}{864 \left[1 + 2\, \textrm{cos} \left(\frac{2}{3} \pi (\frac{\theta}{2 \pi} - 1) \right) \right]^3} \Bigg[88 \sqrt{3} \, \textrm{cos} \left(\frac{2}{3} \pi (\frac{\theta}{2 \pi} - 1) \right) + 20 \sqrt{3} \, \textrm{cos} \left(\frac{4}{3} \pi (\frac{\theta}{2 \pi} - 1) \right) \nonumber\\
%  {} & \qquad \qquad \qquad \qquad \qquad \qquad \qquad - 96 \, \textrm{sin} \left(\frac{2}{3} \pi (\frac{\theta}{2 \pi} - 1) \right) - 60 \, \textrm{sin} \left(\frac{4}{3} \pi (\frac{\theta}{2 \pi} - 1) \right) \nonumber\\
%  {} & \qquad \qquad \qquad \qquad \qquad \qquad \qquad - 27 \, \textrm{tan} \left(\frac{1}{6} \pi (\frac{\theta}{2 \pi} - 1) \right) \left(2 + \textrm{sec}^2 \left(\frac{1}{6} \pi (\frac{\theta}{2 \pi} - 1) \right) \right) + 102 \sqrt{3} \Bigg]\, .
%\end{align}
Again, one needs some intermediate steps:
\begin{align}
  \int_0^\infty dz\, \frac{z (z-1)}{(1+z) (z^3-1)} & = \frac{\pi \left[\sqrt{3} \, \textrm{cos} \left(\frac{\pi}{6} \left(\frac{\theta}{2\pi} - 1 \right) \right) - 3\, \textrm{sin} \left(\frac{\pi}{6} \left(\frac{\theta}{2\pi} - 1 \right) \right) \right]}{3 \left[\textrm{cos} \left(\frac{\pi}{6} \left(\frac{\theta}{2\pi} - 1 \right) \right) + \textrm{cos} \left(\frac{\pi}{2} \left(\frac{\theta}{2\pi} - 1 \right) \right) + \textrm{cos} \left(\frac{5 \pi}{6} \left(\frac{\theta}{2\pi} - 1 \right) \right) \right]}\, ,\\
  \int_0^\infty dz\, \frac{z\, \textrm{log}\, z\, (z-1)}{(1+z) (z^3-1)} & = \frac{\pi^2}{18 \left[\textrm{cos} \left(\frac{1}{6} \pi (\frac{\theta}{2\pi} - 1) \right) + \textrm{cos} \left(\frac{1}{2} \pi (\frac{\theta}{2\pi} - 1) \right) + \textrm{cos} \left(\frac{5}{6} \pi (\frac{\theta}{2\pi} - 1) \right) \right]^2} \\
  {} & \quad \cdot \Bigg[ - 6\, \textrm{cos} \left(\frac{1}{3} \pi (\frac{\theta}{2\pi} - 1) \right) - 6 \, \textrm{cos} \left(\frac{2}{3} \pi (\frac{\theta}{2\pi} - 1) \right) + 6\, \textrm{cos} \left( \pi (\frac{\theta}{2\pi} - 1) \right) \nonumber\\
  {} & \quad \quad + 2 \sqrt{3}\, \textrm{sin} \left(\frac{1}{3} \pi (\frac{\theta}{2\pi} - 1) \right) + 4 \sqrt{3}\, \textrm{sin} \left(\frac{2}{3} \pi (\frac{\theta}{2\pi} - 1) \right) + 2 \sqrt{3}\, \textrm{sin} \left( \pi (\frac{\theta}{2\pi} - 1) \right) - 3 \Bigg]\, . \nonumber
\end{align}

\section{Examples of Thermal Corrections to R\'enyi Entropies}
In this appendix we summarize the thermal corrections to the $n$th R\'enyi entropy for the conformally coupled scalar.
The R\'enyi entropy is calculated with respect to a cap of opening angle $2\theta$ on $S^{d-1}$ for small values of $d$ and $n$.
% few lowest $d$'s ($d > 2$) and $n$'s. 
Define the coefficient $f(\theta)$ such that  $\delta S_n = S_n (T) - S_n (0)$ has the form
\be
  \delta S_n = f(\theta) \, e^{-\beta \Delta / R} + o(e^{-\beta \Delta/R}) \, ,
\ee
where $\Delta = \frac{d-2}{2}$ is the scaling dimension of the free scalar and $R$ is the radius of $S^{d-1}$. 
The following tables give the form of $f(\theta)$.  (We also give results for the entanglement entropy, denoted EE.)
\flushleft For $(2+1)$ dimensions:
\begin{displaymath}
\begin{array}{c|c}
  {\rm EE} & \frac{\pi}{2} \, \textrm{tan} \left(\frac{\theta}{2} \right) \\
  \hline
  n=2 & \frac{2 \theta}{\pi} \\
  \hline
  n=3 & \sqrt{3} \, \textrm{sin} \left(\frac{\theta}{3} \right)
\end{array}
\end{displaymath}

\flushleft For $(3+1)$ dimensions:
\begin{displaymath}
\begin{array}{c|c}
 {\rm EE}  & 2-2\, \theta\, \textrm{cot}(\theta) \\
  \hline
  n=2 & 1-\textrm{cos}(\theta) \\
  \hline
  n=3 & \frac{4}{3} \left[2 + \textrm{cos} \left(\frac{2 \theta}{3} \right) \right] \, \textrm{sin}^2 \left(\frac{\theta}{3}\right)
\end{array}
\end{displaymath}

\flushleft For $(4+1)$ dimensions:
\begin{displaymath}
\begin{array}{c|c}
 {\rm EE}  & 3 \pi \, \textrm{csc}(\theta)\, \textrm{sin}^4 \left(\frac{\theta}{2} \right) \\
  \hline
  n=2 & \frac{1}{\pi} \left[2 \theta - \textrm{sin} (2 \theta) \right] \\
  \hline
  n=3 & \frac{1}{6\sqrt{3}} \left[51 + 44 \, \textrm{cos} \left(\frac{2 \theta}{3} \right) + 10\, \textrm{cos} \left(\frac{4 \theta}{3} \right) \right]\, \textrm{sin}^3 \left(\frac{\theta}{3} \right)
\end{array}
\end{displaymath}

\flushleft For $(5+1)$ dimensions:
\begin{displaymath}
\begin{array}{c|c}
  {\rm EE}  & \frac{2}{3} \left[5 + \textrm{cos} (2 \theta) - 6\, \theta\, \textrm{cot}(\theta) \right] \\
  \hline
  n=2 & 2 \left[2+\textrm{cos}(\theta) \right] \, \textrm{sin}^4 \left(\frac{\theta}{2} \right) \\
  \hline
  n=3 & \frac{16}{81} \left[50 + 60 \, \textrm{cos} \left(\frac{2 \theta}{3} \right) + 21\, \textrm{cos} \left(\frac{4 \theta}{3} \right) + 4\, \textrm{cos} (2 \theta) \right]\, \textrm{sin}^4 \left(\frac{\theta}{3} \right)
\end{array}
\end{displaymath}

\bibliographystyle{utphys}
\bibliography{RenyiPaper}

\end{document}